\documentclass[aps,prd,twocolumn,superscriptaddress,showpacs]{revtex4}
\usepackage{graphicx}
\usepackage{dcolumn}
\usepackage{bm}
\usepackage{natbib}

\newcommand{\apjl}{{Astrophys.~J.~Lett.}}
\newcommand{\apjs}{{Astrophys.~J.~Supp.}}
\newcommand{\aj}{{Astron.~J.}}
\newcommand{\mnras}{{Mon.~Not.~R.~Astron.~Soc.}}

\newcommand{\nhat}{\hat{\bf n}}
\newcommand{\lmax}{\ell_{\rm max}}
\newcommand{\rmin}{(T/S)_{\rm min}}
\newcommand{\threej}[6]{\left(\begin{array}{ccc} #1 & #2 & #3 \\ #4 & #5 & #6 \end{array}\right)}
\newcommand{\nh}{n_{\rm H}}

\newcommand{\WMAP}{{\slshape WMAP}}

\newcommand{\muk}{\,\mu{\rm K}}
\newcommand{\muks}{\,\mu{\rm K}^2}
\newcommand{\kpc}{\,{\rm kpc}}
\newcommand{\Mpc}{\,{\rm Mpc}}
\newcommand{\cm}{\,{\rm cm}}
\newcommand{\msun}{M_\odot}
\newcommand{\sr}{\,{\rm sr}}

\newcommand{\Tr}{{\rm Tr}\,}

\begin{document}

\title{CMB $B$-mode polarization from Thomson scattering in the local universe}

\author{Christopher M. Hirata}
\email{chirata@princeton.edu}
\affiliation{Department of Physics, Jadwin Hall, Princeton University, 
  Princeton, NJ 08544, USA}

\author{Abraham Loeb}
\affiliation{Harvard-Smithsonian Center for Astrophysics, MS-51, 60 Garden Street,
  Cambridge, MA 02138, USA}

\author{Niayesh Afshordi}
\affiliation{Harvard-Smithsonian Center for Astrophysics, MS-51, 60 Garden Street,
  Cambridge, MA 02138, USA}

\date{January 10, 2005}
% ??? Replace date

\begin{abstract}
The polarization of the cosmic microwave background (CMB) is widely recognized as a potential source of information about primordial 
gravitational waves.  The gravitational wave contribution can be separated from the dominant CMB polarization created by density 
perturbations at the times of recombination and reionization because it generates both $E$ and $B$ polarization modes, whereas the 
density perturbations create only $E$ polarization.  The limits of our ability to measure gravitational waves are thus determined by 
statistical and systematic errors from CMB experiments, foregrounds, and nonlinear evolution effects such as gravitational lensing of 
the CMB.  Usually it is assumed that most foregrounds can be removed because of their frequency dependence, however Thomson scattering 
of the CMB quadrupole by electrons in the Galaxy or nearby structures shares the blackbody frequency dependence of the CMB.  If the 
optical depth from these nearby electrons is anisotropic, the polarization generated can include $B$ modes even if no tensor
perturbations are present.  We estimate this effect for the Galactic disk and nearby extragalactic structures, and find that it 
contributes to the $B$ polarization at the level of $\sim (1$--$2)\times 10^{-4}\mu$K per logarithmic interval in multipole $\ell$ for 
$\ell<30$.  This is well below the detectability level even for a future CMB polarization satellite and hence is negligible.  
Depending on its structure and extent, the Galactic corona may be a source of $B$-modes comparable to the residual large-scale lensing 
$B$-mode after the latter has been cleaned using lensing reconstruction techniques.  For an extremely ambitious post-{\slshape Planck} 
CMB experiment, Thomson scattering in the Galactic corona is thus a potential contaminant of the gravitational wave signal; 
conversely, if the other foregrounds can be cleaned out, such an experiment might be able to constrain models of the corona.
\end{abstract}


\maketitle

\section{Introduction}

Recent observations of the cosmic microwave background (CMB) anisotropies have confirmed several of the predictions of the simplest
inflationary models \cite{1981PhRvD..23..347G,1982PhLB..108..389L,1982PhRvL..48.1220A}.  They have (in combination with other
datasets) shown that the universe is close to spatially flat, with $\Omega_{tot}=1.02\pm 0.02$ \cite{2003ApJS..148..175S}; that the
perturbations in the CMB are close to Gaussian \cite{2003ApJS..148..119K}; that the temperature and $E$-polarization power spectra
match those expected from adiabatic initial conditions \cite{2003ApJS..148..161K}; and that the perturbations are close to
scale-invariant with $n_s\approx 1$ and $\alpha_s\approx 0$ \cite{2003ApJS..148..175S}.  Thus there is now interest in testing the
last of the major predictions of inflation:  the existence of a roughly scale-invariant background of tensor (gravitational wave)
perturbations.  The amplitude of the tensor perturbations is typically described with the ratio of the temperature quadrupoles in the
CMB $T/S\equiv C_2^{\rm tensor}/C_2^{\rm scalar}$.  The value of $T/S$ depends on the specific inflationary model through the relation
$T/S=V_\ast/(3.7\times 10^{16}{\rm ~GeV})^4$, where $V_\ast$ is the energy density at the time during inflation when observable scales
in the CMB ($10^{-4}<k<10^{-2} h\;$Mpc$^{-1}$) exited the horizon.  A detection of the tensor perturbations would bolster confidence
in inflation, as well as providing a measurement of $V_\ast$.

If $T/S$ is large, then the tensor perturbations add to the $\ell<100$ temperature power spectrum of the CMB, and the nondetection of
this feature in the {\slshape Wilkinson Microwave Anisotropy Probe} (\WMAP) data sets an upper limit of $T/S<0.7$
\cite{2003ApJS..148....1B}.  However the use of the temperature power spectrum to constrain $T/S$ suffers from the large cosmic
variance error bars on the low CMB multipoles, which impose a fundamental limit on measurements of $T/S$ from $C^{TT}_\ell$.  In
addition, these measurements are model-dependent in the sense that modifications to the scalar power spectrum can produce excess power
at $\ell<100$ without invoking tensors (for example, $T/S$ is partially degenerate with the running of the spectral index $\alpha_s$
in the current data \cite{2003MNRAS.342L..79S}).  An alternative is to use the CMB polarization, which can be decomposed into $E$ and
$B$ modes \cite{1997PhRvD..55.1830Z,1997PhRvD..55.7368K}.  Since the scalar density fluctuations of multipole $\ell$ have parity
$(-1)^\ell$, whereas the $E$ and $B$ polarization modes have parity $(-1)^\ell$ and $-(-1)^\ell$ respectively, it follows that the
scalars can produce only $E$ polarization in linear perturbation theory.  However tensor perturbations consist of both positive and
negative parity modes for every value of $\ell\ge 2$, so the tensors can contribute to both $E$ and $B$.  The measurement of $T/S$
through $B$-mode polarization thus is not limited by the cosmic variance associated with the scalars, and is not degenerate with any
features in the scalar power spectrum; instead it is limited by the observations, foregrounds, and nonlinear processes that can create
$B$-mode polarization from scalar initial conditions.

The dominant nonlinear process that generates $B$-mode polarization in the CMB is gravitational lensing of the $E$
polarization generated during recombination.  On the large angular scales relevant for tensor $B$-mode searches, the
lensing $B$-mode appears as white noise with $C_\ell^{BB}\approx 2.4\times 10^{-6}\muks$; this would set a detectability
limit of $\rmin = 6\times 10^{-5}$ for a full-sky experiment if $\tau=0.17$ and $\rmin = 2\times 10^{-5}$ if
$\tau=0.07$.  (The optical depth to reionization $\tau$ is relevant since this determines the amplitude of large-scale
polarization.)  However, lensing results in higher-order correlations between the $E$ and $B$ polarization modes that
allow one to ``clean'' out the lensing $B$-mode \cite{2002ApJ...574..566H, 2002PhRvL..89a1303K, 2002PhRvL..89a1304K}.  
The amount by which one can clean the $B$-mode using the iterative technique of Ref.~\cite{2003PhRvD..68h3002H} was
computed by Ref.~\cite{2004PhRvD..69d3005S} as a function of the detector noise and beam size; for the most sensitive
experiment considered (0.25~$\mu$K~arcmin noise and 2~arcmin beam), the $B$-mode polarization can be cleaned down to
$C_\ell^{BB}\approx 5.8\times 10^{-8}\muks$, which reduces $\rmin$ by a factor of 40.

The polarized foregrounds contain both Galactic and extragalactic components.  The extragalactic foregrounds such as
point sources are approximately white noise and hence are largest on small angular scales, whereas the Galactic
synchrotron and dust emission are expected to have a significant polarization at low $\ell$.  These can be at least
partially cleaned using their frequency dependence, which differs from the blackbody signature expected from tensors.  
However it is also possible to generate polarization via Thomson scattering of the CMB quadrupole by free electrons in
the Galaxy or other nearby structures.  While this polarization signal is extremely small and is well below the
predicted level of the synchrotron and/or dust foregrounds at all frequencies, it cannot be cleaned using frequency
information since it is a blackbody signal.  It can contain both $E$ and $B$ polarization if the distribution of
scattering electrons is anisotropic, hence it is a potential foreground for tensor $B$-mode searches. The primary
purpose of this paper is to investigate this Thomson polarization from the local universe.

\section{Galactic contribution to polarization}

\subsection{Basic model}

The polarization Stokes parameters ${\bf P}=(Q,U)$ produced by Thomson scattering are most easily determined in the line-of-sight 
formalism \cite{1998PhRvD..57.3290H} as
\begin{eqnarray}
{\bf P}(\nhat)\!\! &=& -{\sqrt{6}\over 10}
\int_0^\infty g(r,\nhat) \sum_{m=-2}^2 [T_{2m}(r\nhat)-\sqrt{6}E_{2m}(r\nhat)]
\nonumber \\ && \times {\bf Y}_{2m}^E(\nhat) dr,
\end{eqnarray}
where
\begin{equation}
T_{2m}(r\nhat)=\int Y^\ast_{2m}(\nhat') T(\nhat')|_{r\nhat} d^2\nhat'
\label{eq:t2m}
\end{equation}
and
\begin{equation}
E_{2m}(r\nhat)=\int {\bf Y}^{E\,\dagger}_{2m}(\nhat'){\bf P}(\nhat')|_{r\nhat} d^2\nhat'
\end{equation}
are the temperature and polarization quadrupole moments of the CMB at position $r\nhat$ relative to the observer, and
\begin{equation}
g(r,\nhat) = e^{-\tau(r,\nhat)} {d\tau(r,\nhat)\over dr}
\end{equation}
is the Thomson visibility function, which depends on the optical depth $\tau(r,\nhat)$ to distance $r$; we have trivially 
generalized from Ref.~\cite{1998PhRvD..57.3290H} to allow $\tau$ to depend on the direction $\nhat$.  Here 
$Y_{\ell m}$ are the spherical harmonics and ${\bf Y}_{\ell m}^{E}$ are the tensor spherical harmonics.  (We normalize these to $\int 
{\bf Y}_{\ell m}^{E\,\dagger}{\bf Y}^E_{\ell m}(\nhat)\,d^2\nhat=1$ over the whole sphere, with the inner product ${\bf 
P}_1^\dagger{\bf P}_2=Q_1^\ast Q_2+U_1^\ast U_2$; this is consistent with Ref.~\cite{1997PhRvD..55.1830Z} and with {\sc CMBFast} 
\cite{1996ApJ...469..437S} but differs by a factor of $\sqrt{2}$ from Ref.~\cite{1997PhRvD..55.7368K}.)  If we break the line-of-sight 
integral into a piece contained within our galaxy ($r<R=100$~kpc) and a piece external to the galaxy ($r>R$), we find to first order 
in $\tau(R,\nhat)$:
\begin{equation}
{\bf P}(\nhat) = [1-\tau(R,\nhat)]{\bf P}_{cosmic}(\nhat) + {\bf P}_{gal}(\nhat),
\end{equation}
where ${\bf P}_{cosmic}(\nhat)$ is the polarization that we compute if we neglect the Galactic Thomson scattering and
\begin{equation}
{\bf P}_{gal}(\nhat) = -{\sqrt{6}\over 10} \tau(R,\nhat) \sum_{m=-2}^2 (T_{2m}-\sqrt{6}E_{2m}){\bf Y}_{2m}^E(\nhat),
\label{eq:pgal}
\end{equation}
(We have assumed that the cosmic quadrupoles $T_{2m}(r\nhat)-\sqrt{6}E_{2m}(r\nhat)$ at position $r\nhat$ inside the Galaxy
can be replaced by their values at the observer, $T_{2m}-\sqrt{6}E_{2m}$, which is a good approximation since the cosmic 
quadrupole is dominated by perturbation modes with wavenumber $k\ll R^{-1}$.)  The optical depth is related to the electron 
distribution through the result $d\tau/dr=n_e\sigma_T/(1+z)$ where $n_e$ is the electron density and $\sigma_T=6.65\times 
10^{-25}\cm^2$ is the Thomson cross section.

If $\tau(R,\nhat)$ were independent of $\nhat$, i.e. if the Galactic optical depth were the same in all directions, then the only 
effect of the Thomson scattering would be to (i) suppress the CMB power spectrum because of the factor of $1-\tau(R,\nhat)$, and (ii) 
generate a small amount of $\ell=2$ $E$-mode through ${\bf P}_{gal}$.  However if $\tau(R,\nhat)$ is anisotropic with fluctuations 
$\delta\tau/\tau\sim 1$, which is the case in the real Galaxy we inhabit, then ${\bf P}_{gal}$ becomes a mixture of $E$ and $B$ modes 
that is generically of order $\tau$ times the cosmic temperature+polarization quadrupole.  In the rest of this paper, we will neglect 
$E_{2m}$ compared to $T_{2m}$ in Eq.~(\ref{eq:pgal}), since even for the \WMAP\ $\tau=0.17$ cosmology we have $C^{EE}_2=0.08\muks$,
whereas the observed temperature quadrupole in the map of Tegmark {\em et~al.} \cite{2003PhRvD..68l3523T} is 
$C^{TT}_2=200\muks$.

Since this temperature quadrupole has already been measured, we can compute the induced polarization if we have a map of the Galactic
electron distribution.  In this paper we compute the polarization ${\bf P}_{gal}$ using the model of Cordes \& Lazio
\cite{2002astro.ph..7156C} for the Galactic distribution of free electrons.  This model is based on pulsar dispersion (DM) and
scattering (SM) measures as well as SMs of other Galactic and extragalactic radio sources, and is appropriate for the electron content 
of the Galactic disk; the predicted optical depth in the direction of the Galactic poles is $4\times 10^{-5}$ (North) or $5\times
10^{-5}$ (South).  This is somewhat larger than the optical depth of $\sim 3.4\times 10^{-5}$ predicted by the older Taylor \& Cordes
model \cite{1993ApJ...411..674T}.

The input data to the Cordes \& Lazio model is sparsely sampled, with a typical density of pulsars at high Galactic latitude
($|b|>20$~deg) of $n\sim 6\sr^{-1}$, and small-scale structure in the electron density map is thus expected to be lost
\cite{2003astro.ph..1598C}.  As a crude estimate, we note that the number of modes per steradian is $\sim\lmax^2/4\pi$; setting this
equal to $n$ implies that structures smaller than $\lmax\sim 9$ should be lost.  (The 14 non-pulsar SMs at high Galactic latitude do
not significantly alter this conclusion.)  The effective $\lmax$ at high Galactic latitude may be even less than this because some of
the pulsars lie within the thick disk and their lines of sight do not probe structure behind them.  Fortunately, this loss of
resolution will not have a large influence on the results of this paper because we will find that the extragalactic contribution
dominates the Galactic thick disk for $\ell\ge 5$.

\subsection{Milky Way corona}

The Cordes \& Lazio model does not include the Galactic corona component.  Indeed, very few constraints on the electron density in the
corona are available, which is unfortunate because the corona may contain a significant fraction of the Milky Way's baryons 
\cite{2004MNRAS.354..477M}.  Models of the Milky Way suggest a baryonic mass in the bulge and disk components of $\sim 5\times 
10^{10}\msun$, whereas the virial mass is estimated at $\sim 10^{12}\msun$ \cite{2002ApJ...573..597K}.  For the currently favored 
baryon fraction $f_b=\Omega_b/\Omega_m=0.17\pm 0.01$ \cite{2003ApJS..148....1B}, this suggests that an additional $\sim 1.2\times 
10^{11}\msun$ of ``missing'' baryons should either
be present inside the Milky Way's virial radius or else must have been ejected.  (Indeed, Ref.~\cite{2002ApJ...573..597K} argued that
at least half of the baryons are missing even though they assumed $f_b=0.1$.  Increasing $f_b$ to 0.17 only makes the problem worse.)  
Assuming that a mass $M_b$ of baryons were ionized and are present in a shell at radius $R$, they present a mean optical depth
\begin{equation}
\bar\tau = \frac{(1-Y/2)\sigma_TM_b}{4\pi m_pR^2}
= 5.9\times 10^{-5} \frac{(M_b/10^{11}\msun)}{(R/100\kpc)^2},
\label{eq:taubar}
\end{equation}
where $Y=0.24$ is the Helium abundance, $m_p$ is the proton mass, and we have neglected the distance from the Earth to the center of 
the Galaxy in comparison with $R$.  The virial radius for the Milky Way is estimated as $r_{\rm 
vir}\sim 260\kpc$ \cite{2002ApJ...573..597K}, so that if a large fraction of the missing baryons are significantly inside the virial 
radius then the corona contribution to the optical depth could be comparable to the disk contribution.  If instead the baryons 
are uniformly distributed in a ball, the radius and optical depth are related to the number density by
\begin{eqnarray}
R &=& 195\kpc\,\left(\frac{M_b/10^{11}\msun}{\nh/10^{-4}\cm^{-3}}\right)^{1/3} \nonumber\\
\tau &=& 4.6\times 10^{-5} \left(\frac{M_b}{10^{11}\msun}\right)^{1/3} \left(\frac{\nh}{10^{-4}\cm^{-3}}\right)^{2/3}.
\label{eq:rtnh}
\end{eqnarray}

One of the available constraints on the optical depth through the corona comes from the dispersion measures (DMs) of pulsars observed
in the Large (LMC) and Small (SMC) Magellanic Clouds.  The Thomson optical depth and the DM along a line of sight are both
proportional to $\int n_e\,dr$, hence they are related by the equation $\tau = 2.05\times 10^{-6}$DM where the DM is in units of
cm$^{-3}$~pc.  The observed DM in the Magellanic Cloud pulsars is greater than that predicted by the Galactic thick disk model
\cite{2002astro.ph..7156C} by 16--91$\cm^{-3}$~pc depending on the pulsar \cite{2003astro.ph..1598C}.  However much of this excess DM
may be internal to the Magellanic Clouds and not part of the Galactic corona.  Cordes \& Lazio \cite{2003astro.ph..1598C} argue for
this interpretation since the DMs of pulsars not in the Magellanic Clouds are only fit if the electron density falls off beyond $\sim
1\kpc$ from the plane of the Galaxy (see their Sec.~3.1.1).  This is also consistent with the factor of $\sim 5$ disparity among DM
excesses in the LMC along lines of sight separated by only a few degrees.  In any case the contribution of the Galactic corona along
the line of sight to the LMC at a distance of $50\kpc$ is limited to DM$\le 16\cm^{-3}$~pc or $\tau\le 3\times 10^{-5}$.  If this
upper limit is typical of lines of sight to $50\kpc$ and we use Eq.~(\ref{eq:taubar}) with $M_b=1.2\times 10^{11}\msun$ and $R=50\kpc$
as an upper limit on the optical depth beyond $50\kpc$, we find that the optical depth through the Milky Way corona is at most
$3\times 10^{-4}$.  If one instead assumes the uniform ball model (Eq.~\ref{eq:rtnh}) for the corona and assigns a baryonic mass
$M_b=1.2\times 10^{11}\msun$, the DM to the LMC provides an upper limit of $\nh<3.2\times 10^{-4}\cm^{-3}$ and $\tau<1.1\times
10^{-4}$.  This upper limit is significantly strengthened if the corona density is assumed to decrease outwards as a power-law more
gradual than $\propto 1/r$ since this places the bulk of the baryons even farther away (if the power law is steeper than $1/r$ the
optical depth diverges at small $r$, and would be cut off by either some core radius or our finite distance from the center of the
Galaxy).

There are other constraints on the density of the Galactic corona.  If the infall velocities of high-velocity clouds are assumed to be
less than their terminal velocity due to drag in the Galactic corona, the density of the corona is limited to $\nh<3\times
10^{-4}\cm^{-3}$ at $100\kpc$ \cite{2001ApJ...555L..95Q}.  Models of the kinematics \cite{1994MNRAS.270..209M} and H$\alpha$ emission
\cite{1996AJ....111.1156W} from the Magellanic Stream argue for $\nh\sim 10^{-4}\cm^{-3}$ and $\sim 5\times 10^{-5}\cm^{-3}$
respectively at radius $50$--$65\kpc$ (but see criticism of these models arguing that survival of the stream then requires
$\nh<10^{-5}\cm^{-3}$ \cite{2000ApJ...529L..81M}).  Interaction of high-velocity clouds with a Galactic corona is also suggested in
order to produce some of the O~{\sc vi} detected in absorption \cite{2003ApJS..146..165S, 2005astro.ph..1061C}.  If the higher values
of $\nh$ \cite{1994MNRAS.270..209M,1996AJ....111.1156W} are correct, and the corona remains this dense out to radii $>100\kpc$, then
it is possible that the optical depth through the halo may be $\sim 10^{-4}$.

In order to assess the $B$-modes produced in the corona, we must remember that it is the {\em fluctuations} in optical depth
$\delta\tau$, rather than the mean optical depth $\bar\tau$, that produces $B$-modes.  We will return to this point in the discussion,
and for now we focus on the smooth Cordes \& Lazio model \cite{2002astro.ph..7156C} for the known population of Galactic electrons.

\section{Power spectrum}
\label{sec:ps}

\begin{figure}
\includegraphics[angle=-90,width=3.2in]{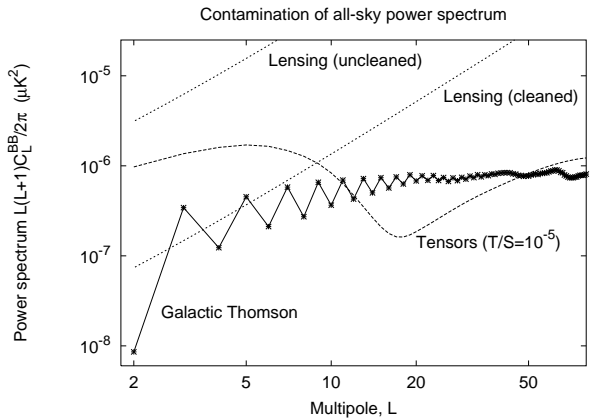}
\caption{\label{fig:allsky}The contributions to the $B$-mode polarization power spectrum.  The white-noise ($C_\ell\propto\ell^2$) 
lines are the lensing signal before cleaning (top line) and after cleaning assuming iterative lensing reconstruction with 0.25 
$\mu$K~arcmin noise and 2~arcmin beam (bottom line).  The dashed line shows the tensor $B$-mode spectrum for $T/S=10^{-5}$.  The 
Galactic Thomson scattering contribution is shown by the solid line with points; note that due to the approximate parity symmetry of 
the Galactic electron distribution, the odd $\ell$ modes have much more power than even $\ell$.}
\end{figure}

We next analyze the $B$-mode power spectrum of the scattered radiation.  In the case of an all-sky experiment, one can simply do the 
spherical harmonic decomposition
\begin{equation}
C^{BB}_\ell({\rm scat}) = {1\over 2\ell+1}\sum_{m=-\ell}^\ell \left|
\int {\bf Y}^{B\,\dagger}_{\ell m}(\nhat){\bf P}(\nhat) d^2\nhat \right|^2,
\label{eq:asps}
\end{equation}
This is shown in Fig.~\ref{fig:allsky}, along with the lensing contribution, and the gravitational waves for $T/S=10^{-5}$ as 
computed by {\sc CMBFast} \cite{1996ApJ...469..437S}.

A realistic low-$\ell$ $B$-mode experiment will exclude the Galactic Plane since this is where the foreground
contamination is the worst.  We would therefore like to understand the $B$-mode contamination on the cut sky.  Here we
will consider cutting out the region within 10 degrees of the Galactic Plane; note that we are cutting this region out
of the final map ${\bf P}_{gal}$, not out of the integration region in Eq.~\ref{eq:t2m}.  Unlike the case of an all-sky
map where Eq.~(\ref{eq:asps}) is the only ``reasonable'' (rotationally invariant) measurement of the $B$-mode power
spectrum, an analysis of the $B$-mode power spectrum on a cut sky requires the choice of an estimator. We choose the
quadratic estimator, which involves the vector ${\bf x}=(Q_1,...Q_N,U_1,...U_N)$ of length $2N$, where $N$ is the
number of pixels.  The covariance matrix ${\bf C}=\langle{\bf xx}^\dagger\rangle$ is then written as
\begin{eqnarray}
{\bf C}_{ij} \!\! &=& {\bf N}_{ij} + \sum_{\ell=2}^{\lmax}\sum_{m=-\ell}^\ell
\Bigl[ C_\ell^{EE} {\bf Y}^E_{\ell m}(\nhat_i) {\bf Y}^{E\,\dagger}_{\ell m}(\nhat_j)
\nonumber \\ && + C_\ell^{BB} {\bf Y}^B_{\ell m}(\nhat_i) {\bf Y}^{B\,\dagger}_{\ell m}(\nhat_j) \Bigr]
\nonumber \\ &=&
{\bf N}_{ij} + \sum_\alpha p_\alpha{\bf C}_{\alpha\,ij},
\label{eq:cij}
\end{eqnarray}
where ${\bf N}$ is the noise covariance matrix, $p_\alpha$ are the parameters of the covariance matrix, and ${\bf C}_\alpha$ are the 
power spectrum templates.  Our objective is to estimate the parameters $p_\alpha$.  The quadratic estimator determines the parameters 
using the relation
\begin{eqnarray}
\hat p_\alpha &=& [{\bf F}^{-1}]_{\alpha\beta}q_\beta,
\nonumber \\
q_\alpha &=& {1\over 2}{\bf x}^\dagger{\bf C}_0^{-1}{\bf C}_\alpha{\bf C}_0^{-1}{\bf x} - {1\over 2}\Tr\left(
   {\bf C}_0^{-1}{\bf C}_\alpha{\bf C}_0^{-1}{\bf N}\right),
\nonumber \\
F_{\alpha\beta} &=& {1\over 2}\Tr\left( {\bf C}_0^{-1}{\bf C}_\alpha{\bf C}_0^{-1}{\bf C}_\beta\right);
\label{eq:quadest}
\end{eqnarray}
here ${\bf F}$ is the Fisher matrix and ${\bf C}_0$ is a positive definite Hermitian weighting matrix.  The quadratic estimator method 
is unbiased regardless of the choice of ${\bf C}_0$ in Eq.~(\ref{eq:quadest}), but is only optimal if ${\bf C}_0$ is the true 
covariance matrix.  (${\bf C}_0$ controls the relative weighting of modes.)  In our case, we will choose ${\bf C}_0$ to contain the 
noise due to lensing after the iterative cleaning, so that it equals ${\bf C}$ if ${\bf N}_{ij}$ is white noise with 0.83$\mu$K~arcmin 
and $C_\ell^{BB}=0$, and $C_\ell^{EE}\gg$noise in Eq.~(\ref{eq:cij}) since the CMB $E$-mode power spectrum is large.

To understand the relationship between this method and the $E/B$ decomposition of Ref.~\cite{2003PhRvD..67b3501B}, note that as 
$C_\ell^{EE}\rightarrow\infty$, any polarization mode ${\bf x}$ that has non-zero inner product with an $E$ mode has ${\bf 
x}^\dagger{\bf C}_0{\bf x}\rightarrow\infty$.  Also, any mode that is orthogonal to all $E$ modes -- i.e. that is a ``pure $B$-mode'' 
in the terminology of Ref.~\cite{2003PhRvD..67b3501B} -- has ${\bf C}_0{\bf x}={\bf Nx}$ and so ${\bf x}$ is an eigenvector of ${\bf 
C}_0^{-1}$ with eigenvalue given by the inverse noise variance per pixel.  Thus the quadratic estimator procedure 
Eq.~(\ref{eq:quadest}) can be thought of as a method for projecting out the pure $E$-modes and ambiguous modes, and then computing the 
power spectrum; the Fisher matrix ${\bf F}$ serves also as the matrix of window functions for the quadratic estimators $q_\alpha$.
(This same type of projection has been suggested by Ref.~\cite{2003PhRvD..68h3509L}.)

\begin{figure}
\includegraphics[angle=-90,width=3.2in]{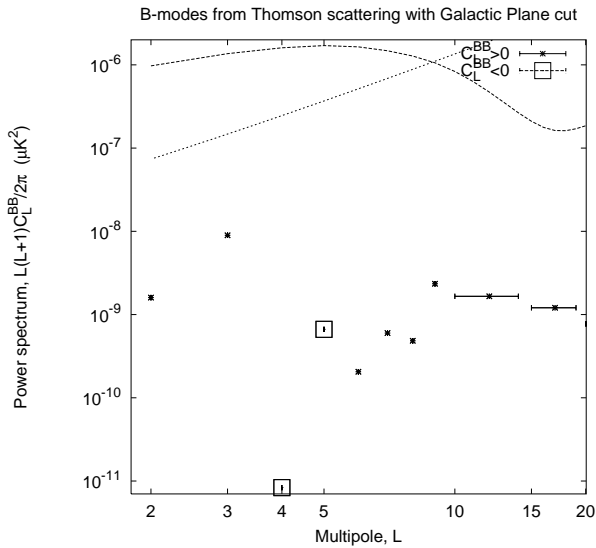}
\caption{\label{fig:b10}The $B$-mode power spectrum on a cut sky (with $|b|<10$~degrees removed) from Thomson
scattering in the Galactic disk, computed with Eq.~(\ref{eq:quadest}).  The dotted line is the cleaned lensing signal assuming 
iterative lensing reconstruction with 0.25 $\mu$K~arcmin noise and 2~arcmin beam, and the dashed line is the tensor contribution
for $T/S=10^{-5}$.  The points show the polarization band powers from Thomson scattering (the open squares represent
negative band powers).  Some power may be missing at $\ell\ge 10$ due to the sparse sampling of the pulsars used to construct the 
electron density model.}
\end{figure}

Numerically, we have evaluated Eq.~(\ref{eq:quadest}) using a weight with $C_\ell^{EE}$ equal to 100 times the noise level for 
$2\le\ell\le\lmax=80$.  We use the implementation of the quadratic estimators described in Ref.~\cite{2004PhRvD..70j3501H} (with no 
preconditioner for the ${\bf C}^{-1}_0$ operations) and the {\sc HEALPix} resolution 5 pixelization system \cite{1999elss.conf...37G}, 
which has 12~288 pixels.  In Fig.~\ref{fig:b10} we have shown the power spectrum of the $B$-mode polarization obtained if we accept 
only the 10~112 pixels at Galactic latitude $|b|>10$~degrees.  We have plugged the resulting polarization data vector ${\bf x}$ of 
length 20~224 into Eq.~(\ref{eq:quadest}), leaving out the $\Tr( {\bf C}^{-1}_0{\bf C}_\alpha{\bf C}^{-1}_0{\bf N})$ term in 
$q_\alpha$ since the Thomson scattering map does not include the noise.  The resulting $C_\ell^{BB}$(scat) represent the additive bias 
to the $B$-mode power spectrum estimators due to the Galactic Thomson scattering component.  (These are negative for $\ell=4,5$; note 
that on a cut sky it is possible for the power spectrum estimator to become negative even without noise subtraction.)

\section{Polarization from scattered Galactic emission}

\begin{figure*}
\includegraphics[angle=-90,width=6.4in]{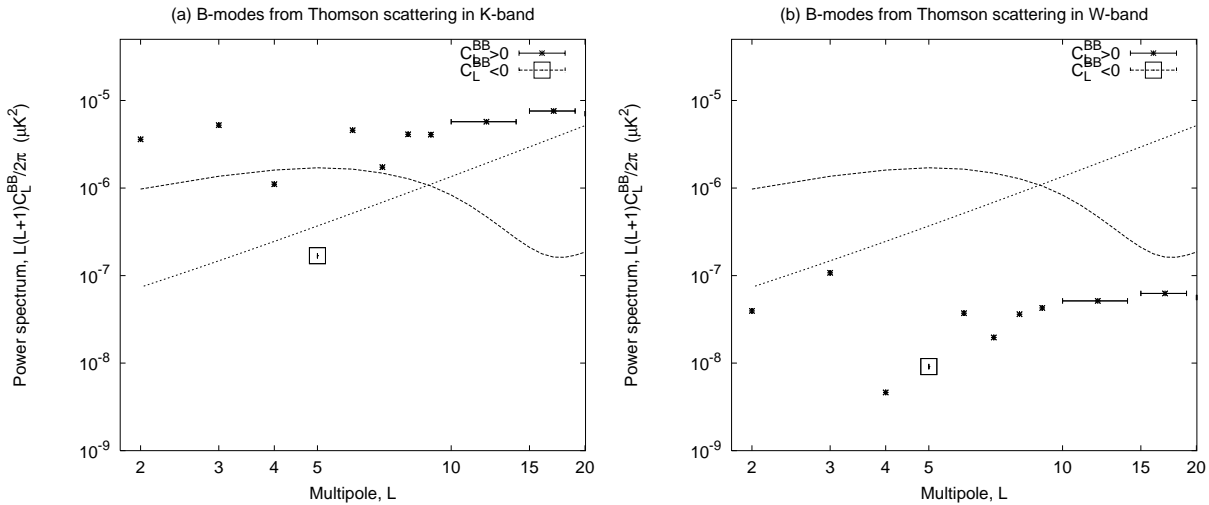}
\caption{\label{fig:b10kw}The $B$-mode power spectrum on a cut sky (with $|b|<10$~degrees removed) from Thomson scattering, in (a) the 
K (23~GHz) band and (b) the W (94~GHz) band.  This figure is the same as Fig.~\ref{fig:b10} except that we have replaced the cosmic 
quadrupole with the raw K and W band quadrupoles observed by \WMAP\ (no frequency cleaning).  Some power may be missing at $\ell\ge 
10$ due to the sparse sampling of the pulsars used to construct the electron density model.}
\end{figure*}

In addition to scattering the CMB quadrupole, Galactic electrons can scatter the microwave foreground emission emitted by the Galaxy 
itself.  In order to compute this effect, one needs not only a model for the distribution of electrons but also for the foreground 
emission, so that one can calculate the quadrupole in Eq.~(\ref{eq:pgal}).  While the foreground emission at frequencies from 
23--94~GHz has been mapped out by \WMAP, the quadrupole $T_{2m}(r\nhat)$ at the scattering electron may differ from the locally 
observed quadrupole because the distance to the electron (of order 1 kpc) is comparable to (and in some cases, greater than) the 
distance to the foreground-emitting regions.  Since \WMAP\ reports the temperature observed in a direction $\nhat$ but cannot specify 
the distance to the emitting region, one needs an additional assumption in order to calculate ${\bf P}_{gal}$.  We will make the 
assumption here that the microwave foreground can be decomposed into a ``local'' contribution from nearby (distances much less than 
the size of the Galaxy) and a ``global'' contribution from distances of order the size of the Galaxy.  The local contribution we 
assume to be plane-parallel; in this case the quadrupole is independent of the position of the scattering electron and we may use the 
observed quadrupole at Earth.  The global contribution to the photon quadrupole at the position of the scattering electron can also be 
taken to equal the foreground quadrupole observed at Earth, since the distance to the scattering electron ($\lesssim 1$~kpc) is small 
compared to the size of the Galaxy.  Being non-axisymmetric, the global contribution can produce both $E$ and $B$ polarization.  
Within this set of assumptions, we can calculate the $B$-modes from Thomson scattering of Galactic foreground emission by repeating 
the procedure of Sec.~\ref{sec:ps} with the uncleaned \WMAP\ maps.  Since these maps contain both foregrounds and CMB, the result -- 
shown in Fig.~\ref{fig:b10kw} -- is the $B$-mode power spectrum from Thomson scattering of both the CMB and the Galactic foreground 
emission.

\section{Extragalactic contribution}

In addition to the Galactic electrons, the CMB quadrupole can also be scattered off of nearby structures in the universe, generating 
$B$-mode polarization.  In this section, we begin by computing the $B$ polarization from structures out to a distance of 80 
$h^{-1}$~Mpc using a model for the local electron distribution.  We then calculate the expected $C_\ell^{BB}$ statistically from the 
power spectrum.  The former method has the advantage of corresponding approximately to the actual realization of the electron 
distribution in the universe, whereas the latter method is better for addressing the contribution from larger distances.

\subsection{Constrained realization method}
\label{ss:cr}

Our first method for estimating the local extragalactic contribution to the $B$-mode polarization is to use the ``constrained 
realization'' $N$-body simulations from Mathis {\em et~al.} \cite{2002MNRAS.333..739M} and assume that the electron density traces the 
dark matter.  These simulations have constrained initial conditions intended to reproduce the structures observed in the local 
universe out to a distance of $\sim 80h^{-1}$~Mpc and on scales larger than $5h^{-1}$~Mpc (on smaller scales, the initial conditions 
are filled in with a Gaussian random field with appropriate power).  Mathis {\em et~al.} \cite{2002MNRAS.333..739M} simulated both a 
$\Lambda$CDM and Einstein-de Sitter cosmology; here we have used the former (with $\Omega_m=0.3$, $H_0=70$~km/s/Mpc, and 
$\sigma_8=0.9$) as it corresponds most closely with the currently favored cosmological parameters.

The perturbation to the optical depth due to nearby extragalactic structures is calculated according to
\begin{equation}
\delta\tau = \sigma_T\int \frac{\bar n_e(z)}{1+z}\delta_e \, dr,
\label{eq:taueg}
\end{equation}
where $\delta_e$ is the fractional electron density perturbation; here we assume $\delta_e=\delta$, which should be
valid except on very small scales where the baryons are segregated from the dark matter.  We have written this as an integral over the 
comoving distance $r$.  The mean electron density assuming complete ionization is
\begin{equation}
\bar n_e = \frac{3\Omega_bH_0^2(1-Y/2)}{8\pi Gm_p}(1+z)^3 = 2.2\times 10^{-7}(1+z)^3\, {\rm cm}^{-3},
\end{equation}
where $G$ is Newton's gravitational constant and the
baryon density $\Omega_bh^2$ is from \WMAP\ \cite{2003ApJS..148..175S}.  Here we have neglected redshift evolution of
$\bar n_e(z)/(1+z)$ and $\delta_e$ since the simulation goes out to a distance of 80$h^{-1}$~Mpc ($z=0.027$); $\bar
n_e(z)/(1+z)$ is only 5.4\%\ greater at $z=0.027$ than today, while $\delta_e$ is slightly less (by 1.3\%\ in the
linear regime).  Equation~(\ref{eq:taueg}) cannot be used directly because the dark matter distribution in the
simulation is represented by individual particles; we have therefore smoothed with a Gaussian of 1$\sigma$ width
0.1$h^{-1}$~Mpc in each dimension.  The Poisson noise at the smoothing scale $k=10h$/Mpc is 4\%\ of the linear matter
power spectrum.  

Physically, the free electrons cannot trace the dark matter down to arbitrarily small scales, and one would expect smoothing on scales 
shorter than the Jeans wavenumber,
\begin{equation}
k_J = \frac{\sqrt{4\pi G\bar\rho(1+\delta)}}{c_s}
= 4.2 h\Mpc^{-1} \sqrt{\frac{10^4\,\rm K}{T_g}(1+\delta)},
\end{equation}
where $c_s$ is the sound speed, $T_g$ is the gas temperature, and we have assumed the sound speed relation $c_s^2=5k_BT/3\mu$ with
mean mass per particle $\mu=0.59m_p$, appropriate for ionized gas with $Y=0.24$.  The intergalactic medium is photoionized so we
expect $T_g\ge 10^4\,$K and hence in regions of average density all structures smaller than $k_J^{-1}\ge 0.2h^{-1}$~Mpc should be
washed out if we consider $\delta_e$ instead of $\delta$.  In this sense our estimate using smoothing at $0.1h^{-1}$~Mpc is
conservative and should overestimate $C_\ell^{BB}$.  Most of the small-scale power is however in overdense regions where $1+\delta\gg
1$, so depending on $T_g$ in these regions the smoothing scale may be different from the typical value of $k_J^{-1}$.  We find
that the $\ell<10$ modes are only slightly affected (by $<20$\% ) if we increase the smoothing length to 0.5$h^{-1}$~Mpc, indicating
that the results are not dominated by the small-scale modes.

Once $\delta\tau$ maps are constructed, these can be fed through Eq.~(\ref{eq:pgal}) to compute the polarization
signal.  The power spectrum on a cut sky (retaining the $|b|>10$~degrees region) can then be determined from a
quadratic estimator, as was done in Sec.~\ref{sec:ps}.

\subsection{Power spectrum method}
\label{ss:psm}

Our second approach to computing $C^{BB}_\ell$ is to estimate the power spectrum of the electron density perturbations,
and then convert this via Eq.~(\ref{eq:pgal}) into a power spectrum for the polarization.  This approach is purely
statistical and does not require any particular realization of the electron distribution.  Its principal disadvantage
is that the sky $\tau(\nhat)$ may look very different from different parts of the universe, and thus the local value of
$C^{\tau\tau}_\ell$ may deviate substantially from the global average computed here.

The perturbation to the Thomson optical depth in direction $\nhat$ is given by Eq.~(\ref{eq:taueg}). We begin with the
case where $\delta_e$ is a Fourier wave, $\delta_e(r,\nhat) = e^{ir{\bf k}\cdot\nhat-\epsilon r}$, where $\epsilon$ is
an infinitesimal parameter that forces convergence of the integral at infinity.  We then have $\delta\tau(\nhat) = \bar
n_e\sigma_T/(\epsilon -ik\mu)$ where $\mu=\hat{\bf k}\cdot\nhat$ is the angle between the wavevector and the line of
sight.  Taking the spherical harmonic transform, we find
\begin{eqnarray}
\delta\tau_{\ell m} \!\! &=& \int_0^{2\pi} d\phi \int_{-1}^1 d\mu\; Y_{\ell m}(\nhat)\tau(\nhat)
\nonumber \\
&=& {i\delta_{m0}\over 2k}\bar n_e\sigma_T\sqrt{4\pi(2\ell+1)} \int_{-1}^1 \frac{P_\ell(\mu)\,d\mu}{\mu+i\epsilon},
\label{eq:dtlm}
\end{eqnarray}
where $P_\ell$ is a Legendre polynomial and we have taken ${\bf k}$ to be along the $z$-axis.  If we define the coefficients
\begin{equation}
a_\ell = \int_{-1}^1 \frac{P_\ell(\mu)\,d\mu}{\mu+i\epsilon},
\end{equation}
then using the recursion relation \cite{1965hmfw.book.....A}
\begin{equation}
\ell P_\ell(\mu)=(2\ell-1)\mu P_{\ell-1}(\mu)-(\ell-1)P_{\ell-2}(\mu),
\end{equation}
and noting that for $\ell>0$ orthogonality implies $\int_{-1}^1 P_\ell(\mu)d\mu=0$, we find $a_\ell = -(\ell-1)a_{\ell-2}/\ell$.  It 
is then easy to show by induction from the initial values $a_0=-i\pi$ and $a_1=2$ that
\begin{equation}
a_\ell = 2^{-\ell}i^{\ell-1}\pi\Gamma(\ell+1)\left[\Gamma\left(\frac{\ell}{2}+1\right)\right]^{-2}.
\label{eq:al}
\end{equation}

To generalize from the case of a single Fourier mode to a random field, we need to incoherently integrate the power spectrum 
$C^{\tau\tau}_\ell$ over all Fourier modes:
\begin{eqnarray}
C^{\tau\tau}_\ell \!\! &=& \int \frac{\sum_{m=-\ell}^\ell |\delta\tau_{\ell m}(\hat{\bf k})|^2}{2\ell+1}
\Delta^2_{\delta_e}(k)\,\frac{dk}{k}
\nonumber \\ &=&
 \frac{\pi^3[\Gamma(\ell+1)]^2(\bar n_e\sigma_T)^2}{4^\ell\{\Gamma[(\ell/2)+1]\}^4}
\int k^{-2}\Delta^2_{\delta_e}(k)\,\frac{dk}{k};
\label{eq:cltt}
\end{eqnarray}
in the last line we have substituted the result of Eq.~(\ref{eq:al}).  (The summation over $m$ is rotationally invariant and so it 
does not matter that we have assumed ${\bf k}$ to be in the $z$-direction.)  This is slightly smaller than the Limber approximation
\begin{eqnarray}
C_{\ell,\,\rm Limber}^{\tau\tau} &=& (\bar n_e\sigma_T)^2 \int 
  r^{-2}\frac{2\pi^2}{k^3}\Delta^2_{\delta_e}(k)|_{k=(\ell+1/2)/r}\,dr
\nonumber \\
&=& \frac{2\pi^2(\bar n_e\sigma_T)^2}{\ell+1/2} \int k^{-2}\Delta^2_{\delta_e}(k)\,\frac{dk}{k}
\end{eqnarray}
by only 5\%\ for $\ell=1$ and 2\%\ for $\ell=2$.

To go from Eq.~(\ref{eq:cltt}) to the $B$-mode power spectrum, we note that from Eq.~(\ref{eq:pgal}), the multipoles of the 
polarization from the local universe are
\begin{eqnarray}
B_{loc,\ell m} &\equiv & \int {\bf Y}^{B\,\dagger}_{\ell m}(\nhat){\bf P}(\nhat)\,d^2\nhat
\nonumber \\
&=& -\frac{\sqrt{6}}{10}\sum_{m'\ell''m''}I^{\ell\ell''}_{mm'm''}T_{2m'}\tau_{\ell''m''},
\label{eq:bloc}
\end{eqnarray}
where we have used the mode coupling integral
\begin{equation}
I^{\ell\ell''}_{mm'm''} = \int {\bf Y}^{B\,\dagger}_{\ell m}(\nhat){\bf Y}^E_{2m'}(\nhat)
   Y_{\ell''m''}(\nhat)\,d^2\nhat.
\end{equation}
Expressing the tensor spherical harmonics in terms of the spin-weighted spherical harmonics,
\begin{eqnarray}
\left[{\bf Y}^E_{\ell m}\right]{_Q} &=& \left[{\bf Y}^B_{\ell m}\right]{_U} = \frac{1}{2}(Y_{\ell m}^2 + Y_{\ell m}^{-2})
{\rm ~~~~~and} \nonumber \\
\left[{\bf Y}^E_{\ell m}\right]{_U} &=& -\left[{\bf Y}^B_{\ell m}\right]{_Q} = \frac{1}{2i}(Y_{\ell m}^2 - Y_{\ell m}^{-2}),
\end{eqnarray}
gives
\begin{eqnarray}
I^{\ell\ell''}_{mm'm''} &=& {1\over 2i}\int(Y_{\ell m}^{-2\ast} Y_{2m'}^2 - Y_{\ell m}^{2\ast} Y_{2m'}^{-2})Y_{\ell''m''}\,d^2\nhat
\nonumber \\
&=& -i(-1)^m \sqrt{\frac{5(2\ell+1)(2\ell''+1)}{4\pi}}
\nonumber \\ && \times \threej{\ell}{\ell''}{2}{2}{0}{-2} \threej{\ell}{\ell''}{2}{-m}{m''}{m'}
\label{eq:ill}
\end{eqnarray}
for $\ell''=\ell\pm 1$ and $I^{\ell\ell''}_{mm'm''}=0$ otherwise by symmetry (for $\ell''-\ell=0$ or $\pm 2$, 
$I^{\ell\ell''}_{mm'm''}=0$ vanishes by parity, for $|\ell''-\ell|>2$ the $3j$ symbols vanish).  Here we have used the three spherical 
harmonic integral \cite{1960amqm.book.....E, 1998PhRvD..57.3290H, 2002PhRvD..66f3008O}.  Now the power spectrum of 
the Thomson-scattered radiation is obtained by taking the mean square value of Eq.~(\ref{eq:bloc}):
\begin{eqnarray}
C^{BB}_\ell({\rm loc}) &=& \frac{3}{50} \sum_{m'\ell''m''} |I^{\ell\ell''}_{mm'm''}|^2 \langle |T_{2m'}\tau_{\ell''m''}|^2 \rangle
\nonumber \\
&=& \frac{3}{50}C_2^{TT} \sum_{m'\ell''m''} |I^{\ell\ell''}_{mm'm''}|^2 C_{\ell''}^{\tau\tau},
\end{eqnarray}
where we may drop cross-terms between different values of $m'$, $\ell''$, and $m''$ since the different multipoles are uncorrelated 
and we have assumed that $\tau$ in the local universe is independent of the CMB quadrupole.  (The value of the sum cannot depend on 
$m$ because of rotational symmetry.)  We may do the summation over $m'$ and 
$m''$ by applying the $3j$ symbol orthogonality relations, yielding (for $\ell''=\ell\pm 1$)
\begin{equation}
\sum_{m'm''} |I^{\ell\ell''}_{mm'm''}|^2 = \frac{5(2\ell''+1)}{4\pi}\threej{\ell}{\ell''}{2}{2}{0}{-2}^2.
\end{equation}
Using the tabulated values of this specific form of the $3j$ symbol \cite{1960amqm.book.....E} gives the final result
\begin{equation}
C^{BB}_\ell({\rm loc}) = \frac{3C_2^{TT}[(\ell+2)C^{\tau\tau}_{\ell-1} + (\ell-1)C^{\tau\tau}_{\ell+1}]}{80\pi(2\ell+1)}.
\label{eq:bb}
\end{equation}

\subsection{Results}

\begin{figure*}
\includegraphics[angle=-90,width=6in]{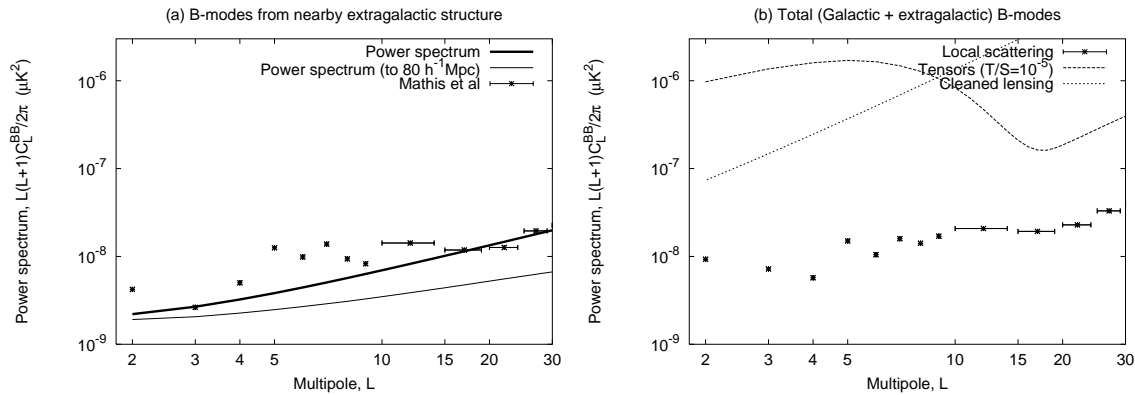}
\caption{\label{fig:extra}The contribution from nearby extragalactic structures to the $B$-modes.  Panel (a) shows the
power spectrum for the constrained realization method of Sec.~\ref{ss:cr} (points), the power spectrum method of
Sec.~\ref{ss:psm} (thick line), and the power spectrum method with the integration cut off at $r_{\rm
max}=80h^{-1}$~Mpc (thin line).  Panel (b) shows the total B-mode from local scattering (points), constructed by taking
the optical depth map from the Galaxy and nearby structures, computing the resulting $C_\ell^{BB}$ using the quadratic
estimator, and then adding the estimated power from $r>80h^{-1}$~Mpc.  For reference, the cleaned-lensing and tensor
curves from Fig.~\ref{fig:b10} are shown.}
\end{figure*}

We can evaluate Eq.~(\ref{eq:bb}) given any power spectrum $\Delta^2_{\delta_e}(k)$, which here we will assume to be
the nonlinear matter power spectrum.  We have computed the spectrum using the Eisenstein \& Hu
\cite{1998ApJ...496..605E} transfer function and the Peacock \& Dodds \cite{1996MNRAS.280L..19P} nonlinear mapping.  
The results from this calculation are shown in the thick line in Fig.~\ref{fig:extra}(a).  For comparison, we have also
shown (thin line) the results obtained by cutting off the integral in Eq.~(\ref{eq:cltt}) at $k_{\rm
min}=(\ell+1/2)/r_{\rm max}$, where $r_{\rm max}=80h^{-1}$~Mpc is the radius to which we integrate in the constrained
realization method.  Within the context of the Limber approximation, this measures the amount of this power that is
recovered by the constrained realization method.  The difference between the thick and thin curves in
Fig.~\ref{fig:extra}(a)  represents the contribution to $C_\ell^{BB}$ coming from structures farther away than
$80h^{-1}$~Mpc; this is 50\% of the power at $\ell=10$, with nearby structures dominating at lower $\ell$ and more
distant structures at higher $\ell$.

Also shown is the power spectrum from the constrained realization method (points in Fig.~\ref{fig:extra}a).  We can see
from the figure that the power spectrum $C_\ell^{BB}$ obtained from this method is greater than the predicted result
out to a distance of $80h^{-1}$~Mpc (thin line).  This indicates that the nearby universe is more inhomogeneous than
average, which is not entirely surprising since we live in an overdense region of the universe and it is these
regions that contribute most of the small-scale power.

Since the Mathis {\em et~al.} \cite{2002MNRAS.333..739M} simulations provide an actual optical depth map, it is possible to construct
an actual realization for the combined Thomson scattering signal from the Milky Way's disk and from nearby extragalactic structures.  
We have computed the power spectrum $C_\ell^{BB}$ of this realization with the quadratic estimator, and added to it the ``missing''
power from $r>80h^{-1}$~Mpc; the result is shown in Fig.~\ref{fig:extra}(b).  This is our final result for the frequency-independent,
Thomson-induced $B$-mode power spectrum from the Galactic disk and nearby extragalactic structures (but excluding the corona).  The
contribution is roughly $(1$--$2)\times 10^{-4}\mu$K per logarithmic interval in $\ell$, which is well below the cleaned lensing
signal even for an optimistic experiment.

We can be more quantitative about the effect on the estimation of the tensor signal $T/S$ by using the tensor signal 
${\bf C}_{T/S}\equiv d{\bf C}/d(T/S)$ as a template in the quadratic estimator instead of the band powers.  In this case we can get 
the spurious contribution to $T/S$ from the equation
$\Delta(T/S) = q_{T/S}/F_{T/S,T/S}$,
where $q_{T/S}$ and $F_{T/S,T/S}$ are computed using Eq.~(\ref{eq:quadest}) with ${\bf C}_{T/S}$.  The map ${\bf x}$ to be inserted 
into the equation for $q_{T/S}$ is the sum of the known Galactic disk and extragalactic ($r<80h^{-1}$Mpc) signals and 
the unknown extragalactic signal from $r>80h^{-1}$Mpc, which we denote ${\bf x}^{(1)}$ and ${\bf x}^{(2)}$ respectively.  We can put 
into the estimator the actual value of ${\bf x}^{(1)}$, whereas we must account for ${\bf x}^{(2)}$ statistically using its covariance 
matrix ${\bf C}^{(2)}$:
\begin{eqnarray}
\Delta(T/S) &=& \frac{\langle q_{T/S}\rangle}{F_{T/S,T/S}} \nonumber \\
&=& (F_{T/S,T/S})^{-1} \Bigl[ \frac{1}{2}{\bf x}^{(1)\dagger}{\bf C}_0^{-1}{\bf C}_{T/S}{\bf C}_0^{-1}{\bf x}^{(1)}
\nonumber \\ &&
  + \frac{1}{2}\Tr({\bf C}_0^{-1}{\bf C}_{T/S}{\bf C}_0^{-1}{\bf C}^{(2)}) \Bigr].
\end{eqnarray}
This equation yields $\Delta(T/S)=1.0\times 10^{-7}$ for the template with optical depth $\tau=0.17$, and $3.6\times 10^{-7}$ for 
$\tau=0.07$.  This is negligible, since even for the optimistic (0.25$\muk\,$arcmin noise, 2~arcmin beam) experiment, the 
Fisher matrix uncertainties in $T/S$ are $6.2\times 10^{-7}$ and $1.7\times 10^{-6}$ for $\tau=0.17$ and $0.07$ respectively.

\section{Discussion}

We have considered the contribution to the CMB $B$-mode polarization from Thomson scattering in the local universe, which is a 
potential contaminant to the gravitational wave signal.  Our findings are that:
\newcounter{disclist}
\begin{list}{\arabic{disclist}.}{\usecounter{disclist}}
\item The Thomson scattering signal from the Galactic disk and from nearby extragalactic structures is well below the weak lensing 
$B$-modes even after the lensing signal is ``cleaned'' out using a lensing reconstruction from a very optimistic experiment.  This 
signal is believed to be negligible.
\item The Galactic disk also Thomson-scatters the foreground radiation emitted by the Milky Way.  At the low multipoles, this 
radiation can exceed the cleaned lensing signal; in the K band it is roughly equal to the uncleaned lensing signal at $\ell=2$--$3$.  
However this signal is strongly frequency-dependent, and in principle can be removed by the same techniques used to clean other 
Galactic foregrounds.
\item Thomson scattering within the hot Galactic corona makes an unknown contribution to the $B$-modes, depending on the flucutations 
in the optical depth through the corona.  The contribution can be roughly estimated based on Eq.~(\ref{eq:bb}):
\begin{equation}
\frac{\ell(\ell+1)}{2\pi}C_\ell^{BB} \sim \frac{3C_2^{TT}}{80\pi}\Delta^2_\tau \sim 2.4\muks\;\Delta^2_\tau;
\end{equation}
this can exceed the cleaned lensing signal if the large angular scale fluctuations in the optical depth through the corona are 
$\Delta_\tau\ge 2\times 10^{-4}$.  Since a mean optical depth through the corona of up to $\bar\tau\sim O(10^{-4})$ appears to be 
reasonable, and there are only very limited constraints on the angular distribution of free electrons in the corona, fluctuations of 
$\Delta_\tau\ge 2\times 10^{-4}$ cannot yet be ruled out.
\end{list}
Of these, the scattering from the Galactic corona is of the most interest.  In principle, detection of this signal would provide 
evidence that the corona contains some of the missing baryons; non-detection of the $B$-mode signal would provide an upper limit on 
the anisotropy of the corona.  However the practical difficulties should not be understated: reaching the ``cleaned lensing'' limit 
of $C_\ell^{BB} = 5.8\times 10^{-8}\muks$ requires a detector sensitivity of $0.25\muk\,$arcmin, two orders of magnitude better than 
the upcoming {\slshape Planck} satellite \footnote{URL:\\ http://www.rssd.esa.int/index.php?project=Planck}.  It is also possible that 
the frequency structure of the polarized foregrounds is sufficiently complicated that extraction of small large-angle $B$-mode signals 
becomes impractical.  Finally it is possible that tensor perturbations, or some other source of frequency-independent $B$-modes such 
as cosmic strings, are present and swamp the Galactic Thomson scattering signal.

From the perspective of CMB gravitational wave searches, the Galactic Thomson scattering signal appears to be negligible for all
except the most futuristic experiments.  Part of the reason for this is the low value of the local quadrupole $C_2^{TT}$, which is
only $\sim 20\%$ of the $\Lambda$CDM expected value; if $\Lambda$CDM is indeed correct, we are in this respect simply lucky.  For the
very futuristic experiments, however, if a frequency-independent $B$-mode signal were observed close to the detectability threshold,
then the interpretation as a gravitational wave source would require that Thomson scattering in the local universe be robustly ruled
out as the source.  This would require significant improvement in our understanding of the Galactic corona.

\acknowledgments

We thank Joseph Taylor and Bob Benjamin for useful conversations about the Galactic electron density, and Nikhil Padmanabhan for his
assistance with the $E/B$ decomposition software.  C.H. is supported by NASA NGT5-50383.  This work was supported in part by NSF
grants AST-0204514, AST-0071019 and NASA grant NAG 5-13292 (for A.L.).  Some of the results in this paper have been derived using the
{\sc HEALPix} \cite{1999elss.conf...37G} package. We acknowledge the use of the Legacy Archive for Microwave Background Data Analysis
(LAMBDA) \footnote{URL: http://lambda.gsfc.nasa.gov/}. Support for LAMBDA is provided by the NASA Office of Space Science.

\end{document}